\documentclass[10pt]{article}
\usepackage[dvips]{graphicx}  

\usepackage{color}

\begin{document}

\renewcommand{\refname}{References}
\renewcommand{\tablename}{\small Table}
\renewcommand{\figurename}{\small Fig.}
\renewcommand{\contentsname}{Contents}

\begin{center}
{\Large\bf Lunar Laser-Ranging Detection of Light-Speed Anisotropy and Gravitational Waves\rule{0pt}{13pt}}\par

\bigskip
Reginald T. Cahill \\ 
{\small\it  School of Chemical and  Physical Sciences, Flinders University,
Adelaide 5001, Australia\rule{0pt}{15pt}}\\
\raisebox{+1pt}{\footnotesize E-mail: Reg.Cahill@flinders.edu.au}
\\ ({\it Progress in Physics}, {\bf 2}, 31-35, 2010)\par

\bigskip

{\small\parbox{11cm}{%
The Apache Point Lunar  Laser-ranging Operation (APOLLO), in NM, can detect photon bounces from  retroreflectors on the moon surface to 0.1ns timing resolution. This facility enables not only the detection of light speed anisotropy, which defines a local preferred frame of reference  - only in that frame is the speed of light isotropic, but also fluctuations/turbulence (gravitational waves) in the flow of the dynamical 3-space relative to local systems/observers.   So the APOLLO  facility can act as an effective ``gravitational wave" detector. A recently published small data set from November 5, 2007, is analysed to characterise both the average anisotropy velocity and the wave/turbulence effects. The results are consistent with some 13 previous detections, with the last and most accurate being from the spacecraft earth-flyby Doppler-shift NASA data.   \rule[0pt]{0pt}{0pt}}}\medskip
\end{center}

\setcounter{section}{0}
\setcounter{equation}{0}
\setcounter{figure}{0}
\setcounter{table}{0}


\section{Introduction}
Light speed anisotropy has  been repeatedly detected over more than 120 years, beginning with the Michelson-Morley experiment in 1887 \cite{MM}. Contrary to the usual claims, that experiment  gave a positive result, and not a null result, and when the data was first analysed, in 2002, using a proper calibration theory for the detector \cite{MMCK,MMC}  an anisotropy  speed, projected onto the plane of the gas-mode interferometer, in excess of 300km/s was obtained.  The problem was that Michelson had used Newtonian physics to calibrate the interferometer. When the effects of a gas in the light path and Lorentz contraction of the arms are taken into account the instrument turns out to be nearly 2000 times less sensitive than Michelson had assumed. In vacuum-mode the Michelson interferometer is totally insensitive to light speed anisotropy, which is why vacuum-mode resonant cavity experiments give a true null result \cite{cavities}.  These experiments demonstrate, in conjunction with the various non-null experiments,  that the Lorentz contraction is a real contraction of physical objects, not that light speed is invariant.  The anisotropy results of Michelson and Morley have been replicated in numerous experiments 
\cite{Miller,Illingworth,Joos,Jaseja,Torr,Krisher,DeWitte,CahillCoax,Munera,CahillFlyby,CahillNASA}, using a variety of different experimental techniques. The most comprehensive early experiment was by Miller\cite{Miller},  and the direction of the anisotropy velocity obtained via his gas-mode Michelson interferometer has been recently confirmed, to within $5^\circ$, using \cite{CahillNASA}   spacecraft earth-flyby  Doppler shift data \cite{And2008}.
The same result is obtained using the range data - from bounce times.  

It is usually argued that light speed anisotropy would be in conflict with the successes of Special Relativity (SR), which supposedly is based upon the invariance of the speed of light. However this claim is false because in SR the space and time coordinates are explicitly chosen to make the speed of light invariant wrt these coordinates.  In a more natural choice of space and time coordinates the speed of light is anisotropic, as discussed in \cite{CahillMink}.  Therein the new  exact mapping between the Einstein-Minkowski coordinates and the natural space and time coordinates is given.    So, rather than being in conflict with SR, the anisotropy experiments have revealed a deeper explanation for SR effects, namely physical consequences of the motion of quantum matter/radiation  wrt a structured and dynamical 3-space.  In 1890 Hertz \cite{Hertz} gave the form for the Maxwell equations for observers in motion wrt the 3-space,  using the more-natural choice of space and time coordinates \cite{CahillMink}. Other laboratory experimental techniques are being developed, such as the use of a Fresnel-drag anomaly in RF coaxial cables, see Fig.6e in \cite{CahillNASA}.
These experimental results, and others, have lead to a new theory of space, and consequently of gravity, namely that space is an observable system with a known and tested dynamical theory, and with gravity an emergent effect from the refraction of quantum matter and EM waves in an inhomogeneous and time-varying 3-space velocity field \cite{Book,Review}.
As well all of these experiments show fluctuation effects, that is, the speed and direction of the anisotropy fluctuates over time   \cite{CahillNASA,Review} - a form of turbulence. These are   ``gravitational waves", and are very much larger than expected from General Relativity (GR). The observational data \cite{CahillNASA} determines that the  solar system is in motion through a dynamical 3-space at an average speed of some 486km/s in the direction RA = 4.29$^h$, Dec = -75$^\circ$, essentially known since Miller's extraordinary experiments in 1925/26 atop M Wilson. This is the motion of the solar system wrt a detected local preferred frame of reference (FoR) - an actual  dynamical and structured system. This FoR is different to and unrelated to the FoR defined by the CMB radiation dipole, see\cite{CahillNASA}.

Here we report an analysis of photon travel time data from the Apache Point Lunar  Laser-ranging Operation (APOLLO)  facility, Murphy {\it et al.}  \cite{Murphy2008}, for photon bounces from retroreflectors on the moon. This experiment is very similar  to the spacecraft Doppler shift observations, and the results are consistent with  the anisotropy results from the above mentioned experiments, though some subtleties are involved, and also the presence of turbulence/ fluctuation effects are evident.

\begin{figure}
\hspace{25mm}\includegraphics[scale=1.5]{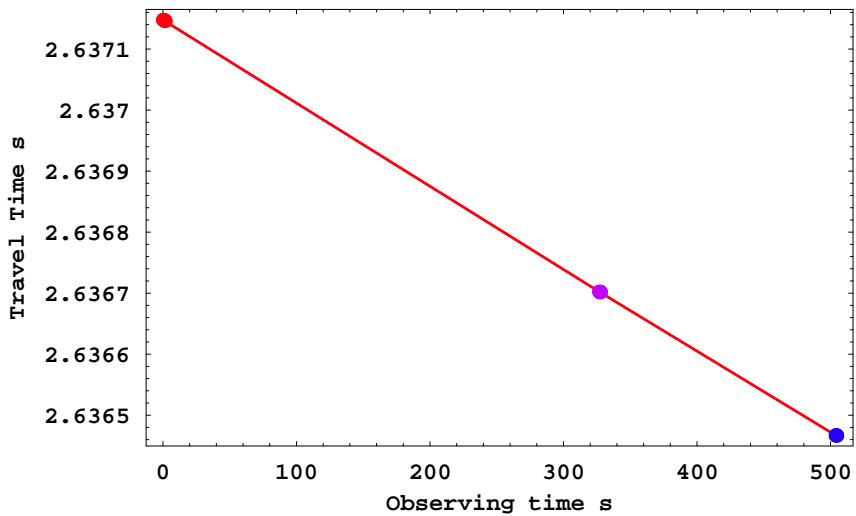}
\vspace{-2mm}\caption{\small{ Total photon travel times, in s, for moon bounces from APO, November 5, 2007, plotted against observing time, in s, after 1st shot at UTC = 0.5444 hrs. Shots 1-5 shown as 1st data point (size of graphic  point unrelated to variation in travel time within each group of shots, typically $\pm$20 ns as shown in Fig.\ref{fig:waveplot}, shots 1100-1104 shown as middle point, and shots 2642-2636 shown in last graphic point. Data from Murphy \cite{Murphy2008}, and tabulated in Gezari \cite{Gezari} (Table 1 therein). Straight line reveals linear time variation of bounce time vs observer time, over the observing period of some 500s.  Data reveals that distance travelled decreased by 204m over  that 500s,  caused mainly by rotation of earth, and using $c$ for light speed.   Data from shots 1000-1004 not used due to possible misprints in \cite{Gezari}.   Expanded data points, after removal of linear trend, and with false zero for 1st shot in each group, shown in Fig.\ref{fig:waveplot}. The timing resolution for each shot is 0.1ns. }
\label{fig:traveltimeplot}}\end{figure}

\section{APOLLO Lunar Ranging Data}

Light pulses are launched from the APOLLO facility, using the 3.5-meter telescope at Apache Point Observatory (APO), NM.  The pulses are reflected by the AP15RR retroreflector,  placed on the moon surface during the Apollo 15 mission, and detected with a time resolution of 0.1ns at the APOLLO facility. The APOLLO facility is designed to study fundamental physics.   Recently Gezari \cite{Gezari}  has published some bounce-time\footnote{Total travel time to moon and back.} data, and performed an analysis of that data. The analysis and results herein are  different from those in\cite{Gezari}, as are the conclusions.  The data is the bounce time recorded from 2036 bounces, beginning at   UTC =  0.54444hrs and ending at UTC=0.55028hrs on November 5, 2007\footnote{The year of the data is not given in \cite{Gezari}, but only in 2007 is the moon in the position reported therein  at these UTC times.}. Only a small subset of the data from these 2036 bounces is reported in \cite{Gezari}, and the bounce times for 15 bounces are shown in Fig.\ref{fig:traveltimeplot}, and grouped into 3 bunches\footnote{An additional 5 shots (shot \#1000-1004) are reported in \cite{Gezari} - but appear to have identical launch and travel times, and so are not used herein.}. The bounce times, at the plot time  resolution,  show a  linear time variation of bounce time vs observer time, presumably mainly caused by changing distance between APO and retroreflector, which is seen to be decreasing over time of observation.  Herein we consider only these bounce times, and not the distance modellings, which are based on the assumption  that the speed of light is invariant, and so at best are pseudo-ranges.

\begin{figure}
\hspace{25mm}\includegraphics[scale=1.5]{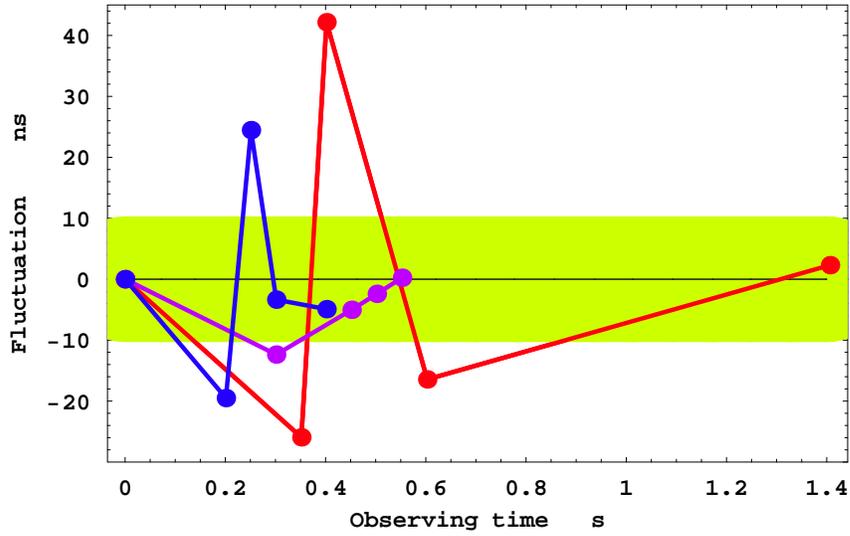}
\vspace{-2mm}\caption{\small{ Fluctuations in bounce time, in ns,  within each group of shots, shown as one data point in Fig.\ref{fig:traveltimeplot}, and plotted against time, in s,  from time of 1st bounce in each group, and after removing the best-fit  linear drift  in  each group, essentially the straight line in Fig.\ref{fig:traveltimeplot}.  The fluctuations are  some $\pm$20 ns. Shaded region shows fluctuation range expected from dynamical 3-space and using spacecraft earth-flyby Doppler-shift  NASA  data \cite{And2008}  for 3-space velocity \cite{CahillNASA}, and using a fluctuation in RA angle of, for example,  3$^\circ$ and a 3-space speed of 490km/s. Fluctuations in only  speed or declination of 3-space produce no measureable effect, because of  orientation of 3-space flow  velocity  to APO-moon direction during these shots.  These fluctuations suggest turbulence or wave effects in the 3-space flow.  These are essentially ``gravitational waves", and have been detected repeatedly since the Michelson-Morley experiment in 1887; see \cite{Review} for plots of that fringe shift  data.}
\label{fig:waveplot}}\end{figure}

Of course one would also expect that the travel times would be affected by the changing orientation of the APO-moon photon propagation directions wrt the light speed anisotropy direction.  However a  bizarre accident of date and timing occurred during these observations.  The direction of the light-speed anisotropy  on November 5 may be estimated from the spacecraft earth-flyby analysis, and from Fig.11 of \cite{CahillNASA} we obtain RA=$5.7^h$, Dec=-75$^\circ$, and with a speed $\approx$ 490 km/s.  And during these APOLLO observations the direction of the photon trajectories was  RA=$11^h 40'$, Dec=$0^\circ 3'$.  Remarkably these two directions are almost at right angles to each other (89.9$^\circ$), and then the speed of 490km/s has a projection onto the photon directions of a mere $v_p=$ 0.8km/s.   

From the bounce times, alone,  it is not possible to extract the anisotropy velocity vector, as the actual distance of the retroreflector is not known.   To do that a detailed modelling of the moon orbit is required, but one in which the invariance of the light speed is not assumed.    In the spacecraft earth-flyby  Doppler shift analysis a similar problem arose, and the resolution is discussed in  \cite{CahillNASA} and \cite{And2008}, and there the asymptotic velocity of motion, wrt the earth, of the spacecraft changed from before to after the flyby, and as well there were various spacecraft with different orbits, and so light-speed anisotropy directional effects  could be extracted.  

\begin{figure}
\hspace{25mm}\includegraphics[scale=1.5]{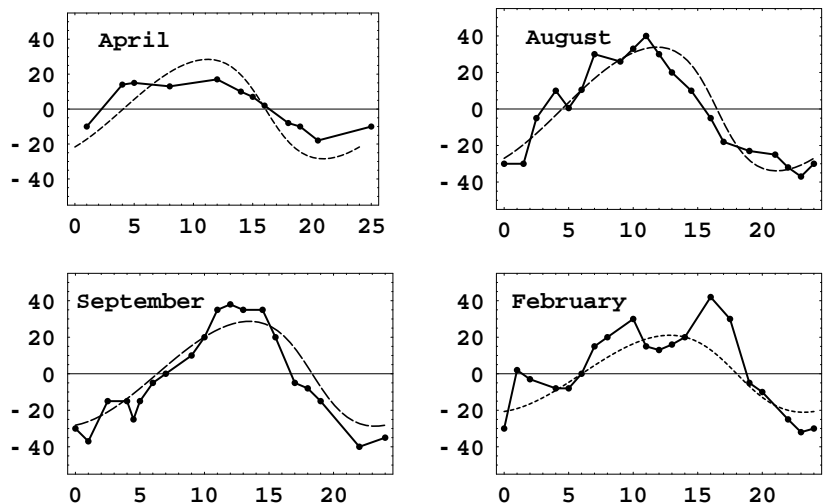}
\vspace{-2mm}\caption{\small{Azimuth, in degrees, of 3-space flow velocity vs local sidereal time, in hrs, detected by Miller \cite{Miller} using a gas-mode Michelson interferometer atop Mt Wilson in 1925/26. Each composite day is a collection of results from various days in each indicated month. In August, for example, the RA for the flow being NS (zero azimuth - here measured from S) is $\approx$ 5 hrs and $\approx$17 hrs. The dotted curves show expected results for the RA, determined in \cite{Book}, for each of these months - these vary due to changing direction of orbital speed of earth and of sun-inflow speed, relative to cosmic speed of solar system, but without wave effects..The data shows considerable fluctuations, at the time resolution of these observations ($\approx$ 1 hr). These fluctuations are  larger than the errors, given as $\pm$ 2.5 $^\circ$ in \cite{Miller}.}
\label{fig:millerplot}}\end{figure}

\section{Bounce-Time Data Analysis}
Herein an analysis of the bounce-time data is carried out to try and characterise the light speed anisotropy velocity.
If the 3-space flow-velocity vector has projection $v_p$ onto the photon  directions, then the round-trip travel time, between co-moving source/reflector/detector system, shows a 2nd order effect in $v_p/c$, see Appendix,
  \begin{equation}
t=\frac{2L}{c}+\frac{L}{c}\frac{v_p^2}{c^2}+......
\label{eqn:triptime}\end{equation}
where $L$ is the actual 3-space distance  travelled.
The last term is the change in net travel time if the  photons have speed $c\pm v_p$, relative to the moving system. There is also a 1st order effect in $v_p/c$ caused by the  motion of the APO site relative to the retroreflector, but this is insignificant, again because of the special orientation circumstance.   These  effects are partially hidden by moon orbit modelling if the invariance of light speed is assumed in that modelling. To observe these $v_p$ effects one would need to model the moon orbit taking into account the various gravity effects, and then observing anomalies in net travel times over numerous  orientations of the APO-moon direction, and sampled over a year of observations.  However a more subtle effect is used now to extract some characteristaion of the anisotropy velocity.  In Fig.\ref{fig:waveplot} we have extracted the travel time variations within each group of 5 shots, by removing a linear drift term, and also using a false zero. We see that the net residual travel times fluctuate by some $\pm20$ns.  Such fluctuations are expected,  because of the 3-space wave/turbulence  effects that have been detected many times, although typically with much longer resolution times.  These fluctuations arise from changes in the 3-space velocity, which means fluctuations in the speed, RA and Dec.  Changes in speed and declination happen to  produce insignificant effects for the present data, because of the special orientation situation noted above, but changes in RA do produce an effect.   In Fig.\ref{fig:waveplot} the shaded region shows the variations  of 20ns (plotted as $\pm10$ns because of false zero) caused by a change in RA direction of $3^\circ$.   This assumes a 3-space speed of 490km/s.   Fig.\ref{fig:millerplot} shows  fluctuations in RA in the anisotropy vector from the Miller experiment \cite{Miller}.  We see fluctuations of some  $\pm2$hrs in RA  ($\equiv  \pm 7.3^\circ$ at Dec =-76$^\circ$), observed with a timing resolution of an hour or so.      So this fluctuation analysis appear to confirm the anisotropy velocity extracted from the earth-flyby Doppler-shift NASA data.  However anisotropy observations have never been made over time intervals of the order of  1sec, as in Fig.\ref{fig:waveplot}, although the new 1st order in $v_p/c$ coaxial cable RF gravitational wave detector detector under construction can collect data at that resolution.

\section{Conclusions}

The APOLLO lunar laser-ranging facility offers significant potential for observing not only the light speed anisotropy effect, which has been detected repeatedly since 1887, with the best results from the spacecraft earth-flyby Doppler-shift NASA data,  but also wave/turbulence effects that have also been repeatedly detected, as has been recently reported,  and which are usually known as ``gravitational waves"\footnote{It may be shown that a dynamical 3-space velocity field may be mapped into a non-flat spacetime metric $g_{\mu\nu}$ formalism, in that both produce the same matter acceleration, but that metric does not satisfy the GR equations \cite{Book, Review}}. These wave effects are much larger than those putatively suggested within GR.  Both the anisotropy effect and its fluctuations show that a dynamical and structured 3-space exists, but which has been missed because of  two accidents in the development of physics, (i) that the Michelson interferometer is very insensitive to light speed anisotropy, and so the original small fringe shifts were incorrectly taken as a ``null effect", (ii)  this in turn lead to the development of the 1905 Special Relativity formalism, in which the speed of light was forced to be invariant, by a peculiar choice of space and time coordinates, which together formed the spacetime construct.  Maxwell's EM equations use these coordinates, but  Hertz as early as 1890 gave the more transparent form which use more natural space and time coordinates, and which explicitly takes account of the  light-speed anisotropy effect, which was of course unknown to Hertz. Hertz had been merely resolving the puzzle as to why Maxwell's equations did not specify a preferred frame of reference effect when computing the speed of light relative to an observer.  In the analysis of the small data set from APOLLO from November 5, 2007, the APO-moon photon direction just happened to be  at 90$^\circ$ to the 3-space velocity vector, but in any case determination, in general,  by APOLLO of that  velocity requires subtle and detailed modelling of the moon orbit,  taking account of  the light speed anisotropy.  Then bounce-time data over a year will show anomalies, because the light speed anisotropy vector  changes due to motion of the earth about the sun, as 1st detected by Miller in 1925/26, and called the ``apex aberration" by Miller, see \cite{CahillNASA}. An analogous technique resolved the earth-flyby spacecraft Doppler-shift anomaly \cite{And2008}. Nevertheless the magnitude of the bounce-time fluctuations can be explained by changes in direction by some $3^\circ$, but only if the light speed anisotropy speed is some 490km/s.  So this is an indirect confirmation of that speed.
Using the APOLLO facility as a gravitational wave detector would not only confirm previous detections, but also provide time resolutions down to a few seconds, as the total travel time of some 2.64s averages the fluctuations over that time interval.  Comparable time resolutions will be possible using a laboratory  RF coaxial cable wave/turbulence detector, for which a prototype has already been successfully operated. Vacuum Michelson interferometers are of course insensitive to both the light speed anisotropy effect and its fluctuations, because  of a subtle cancellation effect - essentially a design flaw in the interferometer, which fortunately   Michelson, Miller and others avoided by using the detector in gas-mode  (air) but without that understanding.

 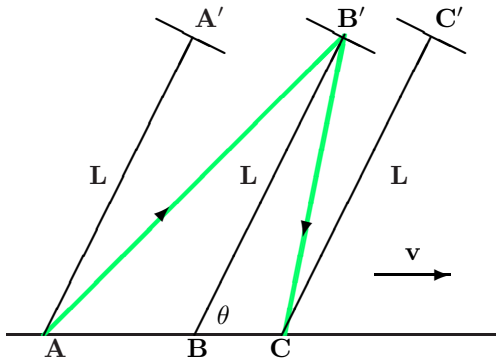
\begin{figure}[ht]
\vspace{15mm}\setlength{\unitlength}{1.0mm}
\hspace{50mm}\begin{picture}(20,30)
\thicklines
\definecolor{green}{rgb}{0, 1,0.4}

\put(-5,0){\line(1,0){65}}
\put(0,-3){{\bf A}}
\put(6,20){{\bf L}}
\put(26,20){{\bf L}}
\put(46,20){{\bf L}}
\put(20,41){{\bf A}$'$}

\color{green}
\put(31.8,0){\line(1,5){8.0}}
\put(32,0){\line(1,5){8.0}}
\put(32.2,0){\line(1,5){8.0}}

\put(0,0){\line(1,1){39.6}}
\put(0.2,0){\line(1,1){39.6}}
\put(-0.2,0){\line(1,1){39.6}}

\color{black}
\put(30,-3){{\bf C}}

\put(19,-3){{\bf B}}
\put(23,+1.5){{$\theta$}}

\put(35,42){\line(2,-1){9}}
\put(39,41){{\bf B}$'$}

\put(44,8){\vector(1,0){10}}

\put(20,0){\line(1,2){20}}

\put(0.0,0){\line(1,2){19.8}}
\put(15,42){\line(2,-1){9}}

\put(31.6,0){\line(1,2){19.8}}
\put(47,42){\line(2,-1){9}}
\put(52,41){{\bf C}$'$}

\put(48,10){{${\bf  v}$}}

\put(14.7,15){\vector(1,1){2}}
\put(34.9,15){\vector(-1,-4){0.5}}

\end{picture}

\vspace{3mm}
\caption{\small{Earth-Moon-Earth photon bounce trajectories in reference frame of 3-space, so speed of light is $c$ in this frame.  Earth (APO)  and Moon (retroreflector)  here taken to have common velocity ${\bf v}$ wrt 3-space.  When APO is at locations A,B,C,  at times $t_A, t_B, t_C,..$ the moon retroreflector is at corresponding locations A$'$, B$'$, C$'$,..  at same respective times $t_A,t_B,t_C,..$ Earth-Moon separation distance $L$,  at same times, has angle $\theta$ wrt   velocity  ${\bf v}$, and shown at three successive times: (i) when photon pulse  leaves APO at A (ii) when photon pulse is reflected at retroreflector B', and (iii) when photon pulse  returns to APO at C.  }
\label{fig:Bounce}}
\end{figure}

\section{Appendix}

 Fig.\ref{fig:Bounce}  shows Earth-Moon-Earth photon bounce trajectories in reference frame of 3-space.  
 Define $t_{AB}=t_{B}-t_A$ and  $t_{BC}=t_{C}-t_{B}$. The distance AB is $vt_{AB}$ and distance   BC is $v t_{BC}$.  Total  photon-pulse travel time is  $t_{AC}=t_ {AB}+t_{BC}$. 
 Applying the cosine  theorem to triangles ABB'  and CBB' we obtain
\begin{equation}\label{eqn:EM}
t_{AB}=\frac{v L\cos(\theta)+\sqrt{v^2L^2\cos^2(\theta)+L^2(c^2-v^2)}}{(c^2-v^2)}.
\end{equation}
\begin{equation}\label{eqn:ME}
t_{BC}=\frac{-v L\cos(\theta)+\sqrt{v^2L^2\cos^2(\theta)+L^2(c^2-v^2)}}{(c^2-v^2)}.
\end{equation}
Then to $O(v^2/c^2)$ 
\begin{equation}\label{eqn:EME}
t_{AC}=\frac{2L}{c}+\frac{L v^2(1+ cos^2(\theta))}{c^3}+.....
\end{equation}
However the travel times are measured by a clock, located at the APO,  travelling at speed $v$ wrt the 3-space, and so undergoes a clock-slowdown effect. So $t_{AC}$ in (\ref{eqn:EME}) must be reduced by the factor $\sqrt{1-v^2/c^2}$, giving
\begin{equation}\label{eqn:EMEdilated}
t_{AC}=\frac{2L}{c}+\frac{L v^2  cos^2(\theta)}{c^3}+.....=\frac{2L}{c}+\frac{L v_P^2  }{c^3}+.....
\end{equation}
where $v_P$ is the velocity projected onto $L$.  
  Note that there is no Lorentz contraction of the distance $L$. However if there was a solid rod separating AA$' $ etc,  as in one arm of a Michelson interferometer, then there would be a Lorentz contraction of that rod, and in the above we need to make the replacement  $L\rightarrow L\sqrt{1-v^2 cos^2(\theta)/c^2}$, giving $t_{AC}=2L/c$ to $O(v^2/c^2)$.   And  then there is no dependence of the travel time on orientation or speed $v$ to  $O(v^2/c^2)$.  
  
Applying the above to a  laboratory  vacuum-mode    Michelson interferometer, as in \cite{cavities},  implies that it is unable to detect light-speed anisotropy because of this design flaw.   The ``null" results from such devices are usually incorrectly reported as proof of the invariance of the speed of light in vacuum. This design flaw can be overcome by using a gas  or other dielectric in the light paths, as first reported in  2002 \cite{MMCK}.

\end{document}